\documentclass[12pt]{article}
\usepackage[utf8]{inputenc}
\usepackage[T1]{fontenc}
\usepackage{amsmath}
\usepackage{amsfonts}
\usepackage{graphicx}
\usepackage{graphicx}
\usepackage{float}
\usepackage{comment, url}
\usepackage{xcolor}

\usepackage{longtable}
\usepackage{titlesec}

\usepackage{hyperref}
\usepackage[left=3cm, right=3cm, bottom=2cm, top=1.5cm]{geometry}
\usepackage{amssymb}
\usepackage{amsmath,xspace}
\usepackage{enumitem}
\usepackage{xurl}

\title{
Specification languages for computational laws versus  
basic legal principles}
\author{Petia Guintchev, Joost J. Joosten, Sofia Santiago Fernández,\\ Eric Sancho Adamson, Aleix Solé Sánchez, Marta Soria Heredia}
\date{March, 2025}

\begin{document}

\maketitle

\begin{abstract}
We speak of a \textit{computational law} when that law is intended to be enforced by software through an automated decision-making process. As digital technologies evolve to offer more solutions for public administrations, we see an ever-increasing number of computational laws.
Traditionally, law is written in natural language. Computational laws, however, suffer various complications when written in natural language, such as underspecification and ambiguity which lead to a diversity of possible interpretations to be made by the coder. These could potentially result into an uneven application of the law. 
Thus, resorting to formal languages to write computational laws is tempting. However, writing laws in a formal language leads to further complications, for example, incomprehensibility for non-experts, lack of explicit motivation of the decisions made, or difficulties in retrieving the data leading to the outcome.
In this paper, we investigate how certain legal principles fare in both scenarios: computational law written in natural language or written in formal language. We use a running example from the European Union's road transport regulation to showcase the tensions arising,  and the benefits from each language.\\
\bigskip

\noindent
\end{abstract}

\maketitle

\newpage

\tableofcontents

\section{Introduction}

We increasingly see Automated Decision Making (ADM) applications in modern administrations. Through ADM, public institutions can automatically assign building permits, compute traffic fines, or decide eligibility for social benefit programs. When administrative ADM is deployed, the agent that makes the legal decision is a computer program. However, the law that establishes the provisions allowing for the decision-making process is written in natural language. If this law is to be implemented by a software program, we often witness a problem: natural language is not adequate for describing algorithms. Typically, natural language allows ambiguity, underspecification and other phenomena that do not fare well with programming.

In this scenario, other than natural language, different expression media could be considered to write laws when they are to be implemented by a computer program. Such an approach will solve natural language law problems for the programmer. However,as we will see, the switch to non-natural language law will introduce many other issues. 

In this paper, we will focus only on those laws that are meant to be implemented by a computer program\footnote{For the purpose of this paper one can either take the common-sense notion of computer program or work with the legal definition as, for example, provided in \cite{Directive2009/24}}. These laws are often referred to as \textit{computational laws}. We will study what is a suitable language for computational laws. 

For our study, we consider two possible languages for the law: natural language, on the one hand, and formal language (better described in Section \ref{section:NaturalFormalLang}), on the other hand, as a non-natural language media of expression. Per language, we list a collection of advantages and draw-backs. To guide our listing and assessment, we organize the benefits and challenges along five fundamental legal principles: Legality, Legal Certainty, Prohibition of Arbitrariness, Motivation, and Clarity.

This approach enables us to deepen our investigations and organize our findings. On the one hand, we will investigate how good practices from ADM affect the legal field and, on the other hand, we will investigate how legal principles do shape corresponding ADM.

% This approach enables us to deepen our investigations and organize our findings. On the one hand, we will contribute to ADM looking at the legal field and, on the other hand, we will contribute to law looking at ADM.

% This approach enables us to deepen our investigations, on the hand, in the logic and AI domain for legal purposes. 

\subsection{Plan of the paper}

This paper is aimed at both the legal scholar and the working programmer or legal IT scholar. Consequently, these communities will find some comments too elementary while others will be very instructive, at the risk of being quite demanding.

We organize our study as follows. This paper presents vastly simplified working definitions of selected legal principles in Section \ref{section:SomeLegalPrinciples}, with the aim of such definitions being useful to the IT scholar. Next, in Section \ref{section:NaturalFormalLang}, we provide a working definition of formal languages and delve onto the differences between natural and formal languages.  In the same section we showcase an article, Article 6.1.~from  Regulation (EC) No 561/2006 
%of the European Parliament and of the Council of 15 March 2006 on the harmonisation of certain social legislation relating to 
on road transport (\cite{Regulation561/2006}) and we will exhibit a formulation both in natural language and in formal language.
%In this section, we give t natural language interpretation of the legal wording in Art.~6.1.~of Regulation (EC) No 561/2006. 

Sections \ref{section:NLIssues} and \ref{sectio:FLIssues} then highlight issues arising from the interactions between the selected legal principles and Art.~6.1.~when expressed in natural or in  formal language.  Section \ref{section:Conclusions} displays a conclusions table derived from our exploration. In that section we intend to raise awareness, at a glance, on the challenges and benefits presented for both the technical and legal communities.

\subsection{On road transportation}

To better explain the choice of our running example, we briefly refer to the legal context and implications of Art.~6.1.~\footnote{In this paper, we use this fragment of the provision from Regulation (EC) No 561/2006, as a mere example of the difficulties that may arise from \emph{computing} Art.~6.1.~of said Regulation. As such mere example, we do not take into account specifics that appear when the article is regarded in relation to other connected articles and regulatory instruments. The example, however, is also useful to showcase problems present in computing other legal provisions.}. For a start, Art.~6.1.~is intended to be computed by a digital tachograph, a digital measuring tool used in road vehicles. 

Tachographs in road transport are regulated by the European Union (EU) under Regulation (EU) No 165/2014 of the European Parliament and of the Council of 4 February 2014 on tachographs in road transport (\cite{Regulation165/2014}). Under Article 2.(2)(a) of the Regulation, the term \emph{tachograph} or \emph{recording equipment} refers to devices designed to display, record, print, store, and output automatically or semi-automatically details regarding the movement of road transport vehicles, including speed, stops and inactivity. Furthermore, these devices are required to record details of certain periods of activity of the drivers to calculate drivers' rest schedules. These details will ultimately determine the drivers’ and employers' compliance with road safety regulations and labor legislation within the European Economic Area.

The EU legal framework governing road transport safety comprises several regulations, with Regulation (EC) No 561/2006 of the European Parliament and of the Council of 15 March 2006 on the harmonization of certain social legislation relating to road transport playing a pivotal role (\cite{Regulation561/2006}). This Regulation establishes various rules pertaining to drivers' working hours, rest periods, and breaks, which, in practice, can present challenges in terms of compliance and enforcement when calculated by digital tachographs. These complexities have already given rise to legal disputes and judicial decisions in EU Member States, such as Spain, France, and Germany\footnote{We refer here to the Judgements of Spain: Sentence 30/2019, CONTENCIOSO/ADMTVO court. N. 4 of Valladolid; France: Tribunal de Police de Cambrai, 5eme Classe, (28/09/2017), Nº de parque 16141000003, Nº Minos: 00104418170050005, Nº Minute: 18/17, and Germany: Amtsgericht Köln Court of 08/Feb/2019, Judgement Number: 902a OWi-912 Js 1132/17 - 73/17.}.

\subsection{This paper compared to existing literature}

For several decades, logic and AI literature has been comparing law in natural language and law in formal languages, warning about key differences in the intellectual techniques used by the two fields: for example, \textit{reasoning} in logic \emph{versus} \textit{argumentation} in law (see, \emph{e.g.}, \cite{Sergot:1986:BritishAct}, \cite{Sergot:1991:IndianCivilService}, \cite{Merigoux:2023:frenchhousing}, \cite{Prakken:2015:LawLogic}. Technology scholars, as well as legal scholars working on the intersection of logic and law, have widely acknowledged the challenges of formally specifying natural language law, primarily due to the open-texture of most legal terms (for all, see \emph{e.g.}, \cite{Prakken:2015:LawLogic}, \cite{Guitton:2024:OpenTextureLaw}) and of the inherent uncertainty of the future (see, \emph{e.g.}, \cite{Hildebrandt:2020:CodeDrivenLaw}). 

However, to the best of our knowledge, these observations have never been measured and organized in relation to their impact on some fundamental (simplified, for the purposes of our paper) legal principles.\footnote{\cite{Smuha_2024} seems to be a reference that is close in spirit to the current paper.} Furthermore, this paper rescales the problem of administrative ADM. Not only do we reassess the general issues identified by other scholars, but we also examine challenges arising from legal terms (like \emph{driving time}, etc.) that appear clear and unambiguous and that have been explicitly defined by the Regulation itself. As we will note, these terms do not seem to exhibit open-texture, yet are problematic. 
% Finally, we connect these issues to our selection of legal principles.

Legal literature also has extensively analyzed the effects of the deployment of ADM (for all see, \emph{e.g.}, \cite{Soriano:2021:AlgorithDiscrim}, \cite{Widlak:2021:PrinciplesGoogDigitalAdm}), although predominantly, the focus has been on fundamental rights instead of legal principles. However, these analyses have not been systematically structured around the distinction between natural and formal language in the specification of ADM systems. The concept of \textit{code as law} has also been largely explored  in legal scholarship (see, \emph{e.g.}, \cite{Boix:2020:AlgortimoReglamento}). However, legal scholars have been particularly interested in its legality and in the legal requirements such code must meet.  In other words, research has examined legal validity criteria of the code when it functions as law. Yet again, to the best of our knowledge, there has been little exploration of what such code should look like and how it should be structured to adhere to existing legal principles while still being logically correct and operable by the ADM system.

The interdisciplinary literature frequently discusses ADM refereed to as AI. However, we stress that this paper does not engage in the recent debate on the distinction between AI-driven ADM and conventional rule-based ADM. We also point out that our working assumption is that ADM is only applied in realms where its use is reasonable. In particular, we exclude scenarios involving discretional powers or domains where numbers cannot describe ontologies. Furthermore, we do not consider ADM based on fancy AI mechanisms, nor do we regard software used for prediction (like \emph{e.g.}, criminal risk scoring). Instead, our focus is on classical computer programs that are applied to arrive at a legal decision in situations where it is not non-sensical to use computer programs. 

We have chosen a particular regulation from traffic law as a typical example. We will see that for either choice (formal language or natural language) there will be benefits and drawbacks and we organise them per principle. We will not position ourselves in favor of a particular choice and just mention the various tensions, so these can be considered in an informed decision.

% We will see that even in such tame environment as traffic law, 

% We will see that even in such tame environment as traffic law, the reasons why one language is suitable for law and the reasons why another language is suitable for computation can be different and, possibly, in tension with one another.

\section{Some legal principles at stake} \label{section:SomeLegalPrinciples}

Legal principles are commonly understood as values deduced from the application of legal rules, contributing to the attainment of fairness (\cite{JRAE})\footnote{This Section does not delve into the definition of \emph{legal principles} nor into the detailed legal discussions regarding the nature and content of the legal principles considered in our paper. In order to avoid such legal discussions, we base our working definitions on two documents of the European Commission for Democracy through the Law (Venice Commission) \cite{VeniceCommissionRuleofLawReport} and \cite{VeniceCommissionRuleofLawChecklist}. The Venice Commission is the Council of Europe's (CoE) advisory body on constitutional matters. The Commission has 61 member States and collaborates with several international organizations, including the EU, to bring their legal and institutional structures in line with the standards of human rights, democracy and the Rule of Law. We also take into account how such principles are widely understood from a comparative perspective, especially within the EU space. For that, we use the European Parliament in-depth analysis \emph{The General Principles of EU Administrative Procedural Law}\cite{EUGeneralPrinciples}, since it lists general principles of EU Law that receive explicit recognition as such from EU Court through the method of \emph{the general principles common to the laws of the Member States}. We also refer to some Spanish Administrative Law principles that have been scholarly consensuated\cite{ManualDerechoAdministrativo}.}.

In the context of the ongoing societal process of digitization across all spheres of life, including ADM (and administrative ADM), public institutions, enterprises, and the citizenry recognize the potential tensions that may arise between the legal framework and the automated implementation of regulations. To address the challenges stemming from legal compliance within a framework of ADM, both soft law and hard law instruments emphasize the importance of safeguarding the rights of citizens and upholding the Rule of Law\footnote{For technical scholars, simply put, \emph{soft law} instruments comprise rules that are not legally binding \emph{per se}, but can inform binding legal rules. Some examples are codes of conduct, recommendations, private quality standards, etc. Conversely, \emph{hard law} instruments are legally binding, regardless of their position within the legal systems hierarchy of legal sources. Some examples are dully ratified international treaties, national constitutions, laws, regulations, etc.}. Such instruments frequently endeavor to establish a list of technology-specific common principles to be respected by digital technologies, drawing inspiration from the applicable legal framework\footnote{To mention only the few most recent legal developments in our geopolitical area, see the Regulation EU laying down harmonised rules on artificial intelligence\cite{Regulation2024/1689}, or the Council of Europe Framework Convention on  Artificial Intelligence, Human Rights, Democracy and the Rule of Law.\cite{CoEFCAI}}. 

However, general legal principles remain applicable even when an ADM process is deployed\footnote{Here we understand \emph{general legal principles} in contraposition to the principles established in soft and hard law instruments that are deemed to guide specifically AI technologies, such as accountability, robustness, human-centeredness, etc.}. These must be observed to ensure the adequacy of the ADM outcome to the legal framework\footnote{For example, Recital 1 of the Preamble of the European Declaration on Digital Rights and Principles for the Digital Decade \cite{EUDeclaratonDigitRights} states that: \emph{1.~The European Union (EU) is a ‘union of values’, as enshrined in Article 2 of the Treaty on European Union, founded on respect for human dignity, freedom, democracy, equality, the rule of law and respect for human rights, including the rights of persons belonging to minorities. Moreover according to the Charter of fundamental rights of the European Union, the EU is founded on the indivisible, universal values of human dignity, freedom, equality and solidarity. The Charter also reaffirms the rights as they result, in particular, from international obligations common to the Member States.} Later, Regulation EU  laying down harmonised rules on artificial intelligence \cite{EUAIAct}, in a similar fashion, in Art. 1.1. establishes that \emph{The purpose of this Regulation is to improve the functioning of the internal market and promote the uptake of human-centric and trustworthy artificial intelligence (AI), while ensuring a high level of protection of health, safety, fundamental rights enshrined in the Charter, including democracy, the rule of law and environmental protection, against the harmful effects of AI systems in the Union and supporting innovation}, thus,  referring explicitly to the EU Charter \cite{EUCharter}.}. However, several fundamental legal principles may exhibit tensions in the context of ADM processes. 

\subsection{Legality}

%\fbox{\fbox { \parbox { .85\linewidth}
\begin{center}
\fbox{\parbox{13cm}{
{\bf Legality: Our simplified working definition}\\
Legality is broadly equivalent to the Supremacy of the Law. State and public authorities must act on the basis and in accordance with the standing legal system.
}}
\end{center}
%}}

Legality, as the foundational pillar of the Rule of Law\footnote{Although the Rule of Law appears in several international treaties and in many domestic legal instruments, it is not a strictly defined concept. Rather, it is understood as a feature of the legal systems that are respectful with the protection and promotion of human rights and of democracy. The Rule of Law concerns all actors in a given legal system and, as a principle, serves to derive obligations towards protecting human rights and democracy, if those are not positively set in the letter of the law.}, mandates that all actions undertaken by the State adhere to and conform to the prevailing legal framework\footnote{Legality encompasses additional dimensions concerning the interplay between international and domestic law, judicial review, and exceptions arising from emergency situations.}.

In the context of our ADM example, essential facets of Legality include the powers granted by the law to public authorities, the legal design of such powers, the procedures required for their exercise, and whether public authorities can delegate public tasks to private entities, that impact on ordinary citizenry, along with the associated legal guarantees.

Additional aspects pertinent to Legality encompass \textit{Transparency} and \textit{Accountability} in legislative procedures and the implementation and enforcement of legal provisions. This raises questions regarding the actors responsible for determining the content of the law, whether the law is implemented consistently, and if the possible sanctions are imposed without incurring discrimination.

\subsection{Legal Certainty}

%\fbox{\fbox { \parbox { .85\linewidth}
\begin{center}
\fbox{\parbox{13cm}{
{\bf Legal Certainty: Our simplified working definition}\\
Legal Certainty is understood as the requirement that those subject to a law must be able to foresee the legal consequences of their actions.
}}
\end{center}
%}}

Legal Certainty is a fundamental legal principle and constitutes another of the pillars of the Rule of Law. The principle of Legal Certainty is guaranteed if its characteristics are complied with by the law. For the purpose of ADM, those also are: foreseeability, stability, and consistency of the law\footnote{According to the \textit{Venice Commission checklist} \cite{VeniceCommissionRuleofLawChecklist}, among the other requirements to be satisfied by the law in order to comply with Legal Certainty are:  accessibility (as publication before entering into force on an official, easily accessible and free of charge platform); prohibition of general retroactivity and of penalties not stated in the law; and the observance of the \textit{res judicata} principle, also known as prohibition of double jeopardy. In the realm of administrative decision-making, the principle of legitimate expectations is also applicable. By the legitimate expectations, the public authorities are required to abide by the law, but also to observe their own prior practices and promises or to dully justify any change of criteria.}. To foster the aforementioned characteristics, accessibility to the legal text is fundamental.

The law is \emph{foreseeable} when its text is formulated clearly and precisely for all subjects whose conduct the law intends to govern. A law with intelligible effects allows the subjects to adjust their behavior and to understand the consequences of non-compliance.

\emph{Stability} is a classic vocation of the law, implying that any law aims to last over time. A stable law addresses and orders a parcel of life in a fashion that is not subscribed to a specific case or time lapse.

\emph{Consistency} of the law is the achievement of uniformity across the application of a given regulatiory instrument. Substantially identical cases will be treated in the same manner and produce the same legal outcome, thus guaranteeing equality.

\subsection{Prohibition of Arbitrariness}

%\fbox{\fbox { \parbox { .85\linewidth}
\begin{center}
\fbox{\parbox{13cm}{
{\bf Prohibition of Arbitrariness: Our simplified working definition}\\
The principle of Prohibition of Arbitrariness interdicts any decision made by public authorities that lacks a sufficient basis, is disproportional or manifestly erroneous, or is deficient in its grounds for motivation, rendering the decision-maker responsible for his or her arbitrary (mis)conduct.
}}
\end{center}
%}}

The principle of Prohibition of Arbitrariness is also an integral part of the Rule of Law and has a direct impact on the decision-making processes of the public authorities\footnote{We also note that the Prohibition of Arbitrariness is enshrined in Art.~41 of the EU Charter of Fundamental Rights, paragraph 1: \textit{1.~Every person has the right to have his or her affairs handled impartially, fairly and within a reasonable time by the institutions, bodies, offices and agencies of the Union}. Such interdiction is common within the domestic jurisdictions and in international legal instruments.}. Public administration must handle citizens' affairs impartially, \emph{i.e.}~without arbitrary action nor unjustified preferential treatment.

This principle is satisfied when administrative processes are handled with due care. Considering all relevant legal and factual elements of a case and pondering all legitimate interests ensures a fair, non-discriminatory, and objective legal outcome.

\subsection{Motivation}

%\fbox{\fbox { \parbox { .85\linewidth}
\begin{center}
\fbox{\parbox{13cm}{
{\bf Principle of Motivation: Our simplified working definition}\\ The Principle of Motivation establishes an obligation to the public administration of giving reasons of the legal bases applied to a case, as well as to the relevance of the factual data leading to any administrative decision.
}}
\end{center}
%}}
The Principle of Motivation, also known as the Principle of Giving Reasons, obliges the public administration to justify the legal grounds applied in an administrative decision and the interaction of such grounds with the relevant facts leading to the decision in question\footnote{This Principle applies to both the Judiciary and the public administration. Attending the scope of our paper, we only refer to the public administration.}. The function of the Principle of Motivation is to enable the affected citizen to comprehend the decision, so s/he can contest it.

Under this principle, the authoring institution becomes responsible when issuing a decision, if such a decision is not well-reasoned, erroneous, malevolent, or produces absurd results. The decision-maker is burdened with accounting for the correct selection of the legal arguments and, in most cases, of the adequacy of the factual data leading to the decision. 

Thus, individuals and collectives affected by a faulty decision resulting in unjustified wrongs are granted the opportunity to receive effective redress by the decision-making institution\footnote{Motivation oftentimes goes hand in hand with the Principle of Accountability. In that sense, other that the affected individual that contests a given decision, valid tools for achieving public administration accountability, as well, and to force a quality motivation satisfaction are civil society and independent institutions' monitoring and oversight, as well as judicial control over public administration decisions.}.

\subsection{Clarity}

%\fbox{\fbox { \parbox { .85\linewidth}
\begin{center}
\fbox{\parbox{13cm}{
{\bf Principle of Clarity: Our simplified working definition}\\
The Principle of Clarity implies that the executive branch must use accessible, easy to understand language both in the regulations it produces and in the acts funded on the legal framework.
}}
\end{center}
%}}

The Principle of Clarity is an extension of the Principle of Legal Certainty, and is most often found in domestic law. It is part of the principles guiding the rationalization of the administrative procedures and can also be considered enshrined in the right to good administration of Art.~41 of the EU Charter of Fundamental Rights (\cite{EUCharter})\footnote{In our paper we are considering this principle of the Spanish Administrative Law. Alongside the Principle of Clarity, we find  to ofentimes cited the Principle of Simplification of the administrative processes and of the burdens imposed on the citizenry, or the Principle of Proximity to the citizens.}.

Clarity directs the drafting of legal instruments by the public administration, which requires that the text is easily comprehensible to \emph{the common citizen}. Consequently, it prohibits unnecessary legal jargon to reinforce Legal Certainty.

The same standard applies to other acts produced by the public administration. Administrative processes and procedures should be comprehensible in their design and development. Similarly, the outcomes of such processes and procedures must be intelligible to the citizenry.

\subsection{On the choice of these principles}

We have selected several aspects of commonly accepted legal principles, as recognized by scholarly standards, corresponding to what we consider to be the primary issues that Regulation 561/2006 (\cite{Regulation561/2006}) may present in relation to Art.~6.1. As a reminder, Regulation 561/2006 establishes conditions for inland road transport to improve the working conditions and safety of drivers. To verify compliance with Regulation 561/2006, Regulation 165/2014 (\cite{Regulation165/2014}) sets forth the rules for the implementation of recording equipment in road transport. Such equipment primarily consists of digital tachographs, although, in some emergency cases, manual recording sheets may be used.

We have chosen Art.~6.1.~of Regulation 561/2006 due to its brevity and simple, straightforward wording. This selection serves illustrative purposes, as we acknowledge that it necessarily interacts in greater depth with the connected provisions of Regulation 561/2006 and the broader legal framework related to road transport. However, even simply considering this point of the article, its natural language wording and its formal language specification highlight challenges and opportunities in its interaction with our chosen legal principles in the setting of ADM.

\section{Natural vs.~Formal Languages} \label{section:NaturalFormalLang}

In the context of formal versus natural legal language, a \emph{specification} should relate our behavior and evaluate whether our behavior complies with legal provisions or not. The specification should, thus, tell us precisely in which situations the behavior is legal and in which it is illegal. Traditionally, the specification of (il)legal behavior is given in natural language: a legal provision that explains how to go from a set of data describing behavior to the decision of legality or illegality. In the realm of ADM, this decision will be reached through a computer program. It makes sense to use a language that is closer to the language used by computers.

As pointed out before, some scholars would delve into the so-called \textit{code as law} paradigm in the context of ADM (see, \emph{e.g.}, \cite{Lessig2000code}, \cite{Hassan2017expansion}). In this paradigm the computer program directly \textit{is} the law. 

This paper does not study this viewpoint of \emph{code as law} since it comes with many legal complications. For one, should a legal instrument or provision really correspond to one exact programming language (C++, Java, Pascal, etc.)?  In this paper, we rather consider the possibility where legal provisions for ADM are written in a \textit{formal language} and not directly as code. By and large, formal languages are mathematical languages that mathematically describe the behavior of the reasoning in ADM that leads from data to outcome or decision. 

On the \emph{syntactic level}, natural language is governed by natural language grammar which allows us to tell, for example, when an English sentence is a proper one and when it is not. Although intensively studied, there is some leeway in telling grammatical and nongrammatical sentences apart. Formal languages have a mathematical definition that indicates exactly which strings of symbols are well-formed formulas and which strings of symbols are not.

For natural language, \emph{semantics} are understood as the reference of our sentences to the real world. Naturally and fortunately, this leaves room for much vagueness, ambiguity, undrspecification and inter-subjectivity\footnote{We point out that, \emph{vaguness} usually refers to imprecise concepts, while \emph{ambiguity} allows for a multiple readings of the wording.}: words like, \emph{sunny}, \emph{nice}, \emph{cloud}, \emph{fast} are inherently open-ended and lean on informal, changing conventions. 

With this respect, the semantics of formal languages are relatively poor: formulas obtain meaning exclusively inside a formally defined semantic framework. For example, sentences in the formal language of arithmetic can only be evaluated in a formal and mathematical framework like that of the natural numbers or similar. As such, sentences in a formal language are unambiguous: they mean exactly one thing. However, this meaning only concerns formal semantics.

What is the usefulness of formal languages to legislators? Formal languages speak about formal semantics as a mathematical structure and legal provisions typically speak about an actor's behavior that can occur in real life. The relation between formal semantics and real life is given by what is called \textit{ontologies}: the relevant entities that a given language speaks of. A typical example of an ontology is \emph{speed}: it is something that the law can speak of and that can be linked to a formal ontology through an observable quantity that is translated to a number\footnote{Often, legal provisions employ ontologies that are harder to relate to observables. An example that we shall see is \textit{driving time}. Somehow, the notion of driving time should be linked to observable quantities like speed, motor activity, driver card insertion (yes/no), etc.}. Moreover, numbers can substitute formal semantics as we have seen above. In short, formal languages can only speak about formal semantics. 

Suppose formal languages should be used to specify whether an actor’s behavior is legal or not. In that case, this behavior should first be described in terms of the objects in the formal semantics: typically numbers. Only in situations where the citizen's behavior can be faithfully expressed by numbers or similar it does make sense to apply ADM\footnote{However, we are very well aware that ADM decisions with legal relevance are being issued in realms where such behavior cannot be or is not being faithfully expressed by numbers. Some of these uses have been problematic, raising concerns about errors (\emph{bugs} and biases).}.

\subsection{A natural language specification of Art.~6.1.~}\label{section:natLangSpecArt61}

We will now focus on an article from EU Regulation 561/2006 on road transport that we have chosen as our example.

Art.~6.1.~of the cited Regulation is stated in natural language:
\begin{quote}\sf

Article 6.1.: The daily driving time shall not exceed nine hours. However, the daily driving time may be extended to at most 10 hours not more than twice during the week.

\end{quote}

The natural language wording of Art.~6.1.~appears to be straightforward and easy to understand, but Section \ref{section:NLIssues} shows various subtleties\footnote{Other than the consideration to be made explicit in Section \ref{section:NLIssues}, we emphasize that Art.~6.1.~combines words (nine) and digits (10) to express time periods of the same nature in its wording, which could be viewed as inconsistent in respect to the form if expression used by the Regulation.}; subtleties that if not addressed in text by the law should be addressed by any programmer that is to implement the provision. %Consequently, it should not pose significant challenges when \emph{computabilized} to verify compliance through a digital tachograph.

\subsection{On formal language specifications}

We follow now by focusing on a possible formal specification of Art.~6.1. When formulating the law in a formal language, the first choice is which formal language. There is no canonical choice; rather, various considerations must be taken into account.

First, the formal language should be expressive enough so that the regulation is expressible with respect to the formal semantics/data that represent the behavior of the citizen/actor. However, too much expressiveness of the formal language is bad: it may cause corresponding computations to become intractable or even undecidable (see \emph{e.g.}, \cite{mullerJoosten:2023:modelcheckingPreprint} for a more detailed discussion). Thus, if the formal language is too expressive, then computations may simply require too much time to be executed. Another consideration is that the chosen formal language used to express the law should allow for succinct and readable expressions\footnote{See \cite{BorgesEtAll:2020:DriveOrNotToDrive} and \cite{mullerJoosten:2023:modelcheckingPreprint} for examples of legal specifications that simply become too long in the chosen specification language.}.
Yet another consideration has to do with errors: it is known that virtually all computer programs contain errors, bugs and flaws\footnote{See,\emph{e.g.}, \cite{McConell:2004:CodeComplete}}. So-called \textit{formal methods} allow one to apply mathematical/logical methods to eradicate errors\footnote{See, \emph{e.g.}~\cite{Oregan:2017:GuideFormalMethods}}. Not all programming languages do allow formal methods so it may be good to choose a specification language for the law that is amenable to formal methods for the implementation. See Subsection \ref{section:formalVerification} for more discussion.

Based on the above considerations, with special weight given to formal methods, in this paper we have chosen to show what a formalization of Art.~6.1.~could look like in the formal language \textit{Gallina}. Details are not really important here\footnote{The interested reader can consult, \emph{e.g.} Gallina \cite{GallinaLangManual}. Gallina is the programming language for the Rocq Prover that can be used to prove program correctness. The Coq Proof Assistant has been renamed as \emph{The Rocq Prover}.  The language of the program (not the specification) is OCAML \cite{OCaml}.}. For the sake of this paper, the main reasons to include an example of a legal article written in formal language are:
\begin{itemize}
    \item 
    to convey the idea of exact and delimited syntax;
    \item 
    to show that formal language resembles programming languages;
    \item 
    to show how formal languages work on numbers that is the data that represents behavior of an agent (a driver in this case);
    \item 
    to show how specifications may invoke other specifications giving rise to nested specifications;
    \item 
    to show that formal languages are extremely precise and do not allow for ambiguity regarding their formal semantics;
\item 
to show that quite some knowledge is required so that one can read a formal specification;

    \item 
    to show how all data and computation steps can be endowed with a so-called \textit{type}.
\end{itemize}

The reader who takes these observations at face value can skip the next subsection. However, we do think it is instructive to skim it through. In particular, for the legal scholar the next subsection may convey a certain taste of formal specifications.

An important feature of Gallina is that it workes with so-called \textit{types}. \emph{Type} is a technical term used in the world of formal methods. We will not be too precise in explaining what a type is. One can think of a data-type. Examples of types are: natural numbers, integer numbers, lists of natural numbers, functions from natural numbers to natural numbers, lists of functions from natural numbers to natural numbers, functions from natural numbers to integer numbers, functions from (functions from natural numbers to natural numbers) to integers, etc\footnote{We recall that the natural numbers are constituted by $\{ 0, 1, 2, 3, \ldots\}$ whereas the integer numbers also include the negative numbers: $\{ \ldots -3, -2, -1, 0, 1, 2, 3, \ldots\}$}. 

A type system is a syntactic framework used in formal methods to classify the elements of a system (or program), allowing for automatic checking of certain erroneous behaviors, such as type mismatches or invalid operations (see, \emph{e.g.},  \cite{Pierce:2002:TypesProgLang}).

To avoid possible errors, Gallina (and Rocq) is very strict on types. For example, it will not be possible to directly compare the integer number $2$ to the natural number $3$. We can only say $2<3$ when both numbers $2$ and $3$ are of the same type (for example, both natural numbers). There are famous software bugs where one team used meters and the other feet\footnote{See, \emph{e.g.}, \cite{NASAMarsClimaOrb} or \cite{ESASchiaparelli}}. 

\subsection{A formal language specification of Art.~6.1.}\label{section:formLangSpecArt61}

We will shortly see how Art.~6.1.~could be formalized in Gallina. For our example formalization, we make certain assumptions. First, we assume that we have fixed and defined a certain data-type for time durations. This data type will be denoted by \texttt{time} and is not a part of Gallina; we assume we have defined it. We do not specify further details and \texttt{time} could be related to hours, minutes, or seconds; the underlying data-type could be natural numbers, or even real numbers. In the context of Regulation 561/2006 it makes sense for \texttt{time} to be natural numbers that denote minutes. For the sake of our example, we shall however assume that \texttt{time} are natural numbers that denote hours.

Some further assumptions are that we have already defined/specified some other functions. These other functions have names that are chosen by us and in the example they are \texttt{is\_weeklyDP}, \texttt{is\_leq\_10} and \texttt{is\_gt\_9}. Thus, all of \texttt{is\_weeklyDP}, \texttt{is\_leq\_10} and \texttt{is\_gt\_9} are not part of Gallina syntax and are chosen by us as names for specifications that we assume we have made. Shortly we shall comment more on those. 

Our specification will be a definition of a function that tells us whether a given driving pattern is legal regarding Art.~6.1.~or not. There exists a precisely delimited Gallina syntax to define functions.
In particular, we use the command \texttt{Definition}, which has the following syntactic structure:

\begin{quote}
    \texttt{Definition \textcolor{red}{name} (\textcolor{red}{a} : \textcolor{red}{A}) : \textcolor{red}{B} := \textcolor{red}{body}.} \ \ \ \ \ \ \ \ \ \ (*)
\end{quote}

Here, fixed syntax is shown in black, while parts that the user can modify in order to declare a specific definition are shown in \textcolor{red}{red}. Specifically: 
\begin{itemize} 
\item \texttt{\textcolor{red}{name}} is the function name, 
\item \texttt{\textcolor{red}{a}} is the input, 
\item \texttt{\textcolor{red}{A}} represents the input type, 
\item \texttt{\textcolor{red}{B}} is the output type, and 
\item \texttt{\textcolor{red}{body}} is the body of the function, that is, some code that tells what the computation consists of. 
\end{itemize} 

With all the above in mind we can now give the formal specification of Art.~6.1~in Gallina:

\textbf{Formal specification of Article 6.1.:}
\begin{verbatim}
  Definition article6_1 (w : list time) : bool :=
  is_weeklyDP w ==> all is_leq_10 w && count is_gt_9 w <= 2.
\end{verbatim} 

We stress that these two lines of specification are not the entire specification: they presuppose that we defined at an earlier stage both the data type \texttt{time} and the functions, \texttt{is\_weeklyDP}, \texttt{is\_leq\_10} and \texttt{is\_gt\_9}.

We can parse this formal specification along the format given in (*). Using the syntax from (*), the function formalizing Art.~6.1.~has the following distinct components:
\begin{itemize}
    \item ``\textcolor{red}{\texttt{article6$\_$1}}'' is the function name, so ``\texttt{\textcolor{red}{name}}'' in the general syntax description (*).
    
    \item \textcolor{red}{\texttt{w}} is the input in our formalization of {\texttt{article6$\_$1}} and so would correspond to \texttt{\textcolor{red}{a}} in the general syntax description (*).
    
        \item  \textcolor{red}{\texttt{list time}} is the type of \textcolor{red}{\texttt{w}} hence would correspond to $\textcolor{red}{\texttt{A}}$ in (*). As mentioned above, \texttt{time} is assumed to be a predefined type for time periods. Typically, \texttt{time} is taken to be natural numbers but it can also be another data type like real numbers, rational numbers, or whatever.  In Gallina, \texttt{list} is a type constructor. That is, it takes a type $t$ and then creates a new type \texttt{list} $t$ which  is the type of lists of elements of type $t$. Thus, \texttt{list time} denotes the type of lists of time periods.
    \item \textcolor{red}{\texttt{bool}}, the output type, represents boolean values (\texttt{true} or \texttt{false}).
Thus, \textcolor{red}{\texttt{bool}} would correspond to $\textcolor{red}{\texttt{B}}$ in (*).
    \item \textcolor{red}{\texttt{is\_weeklyDP w ==> all is\_leq\_10 w \&\& count is\_gt\_9 w <= 2}} is the function body corresponding to $\textcolor{red}{\texttt{body}}$ in (*).
\end{itemize}

Here, Art.~6.1.~is formalized in the Gallina language as the function that we define and named \texttt{article6\_1}.  In general, a function is an expression that defines a formal relationship between an input and a uniquely related output. In particular, \texttt{article6\_1} takes a list \texttt{w} of time periods (\texttt{list time}) as input and returns a so-called boolean value (\texttt{bool}, either \texttt{true} or \texttt{false}). The value of \texttt{article6\_1 w} is \texttt{true} (indicating that Art.~6.1.~holds for the driving pattern recorded by \texttt{w}) if and only if the following conditions are met:

\begin{quote}
if ${\tt w}$ is a weekly driving period, then every element of ${\tt w}$ is at most $10$ and there are at most two elements in ${\tt w}$ that are greater than $9$.
\end{quote}

Let us delve just a little bit more into the details of Gallina. Gallina has its own way of parsing code. From primary school we know that $3\times 2 + 1$ stands for $(3\times 2) + 1$ and not for $3\times (2 + 1)$. Likewise, Gallina knows that the provided code should be parsed as 
\begin{verbatim}
  Definition article6_1 (w : (list time)) : bool :=
  ((is_weeklyDP w) ==> (((all is_leq_10) w) && (((count is_gt_9) w) <= 2))).
\end{verbatim}

We declared that \textcolor{red}{\texttt{w}} is a list of driving periods and we assume that we already have defined/specified three other functions. In particular, we have assumed that \textcolor{red}{\texttt{is\_weeklyDP}} is a function that has as input a list of durations (\textcolor{red}{\texttt{w}}) and outputs \texttt{true} only in case that \textcolor{red}{\texttt{w}} has the right format: a sequence of daily driving periods during a week\footnote{See Section \ref{section:NLIssues} for some comments on \emph{week} and \emph{day} versus \emph{daily}.}. 

Furthermore, we assume that we have defined/specified a function \textcolor{red}{\texttt{is\_leq\_10}} whose input is
an element of type \texttt{time} and whose output is a boolean with value \texttt{true} just in case the value of the input is at most 10. Functions that output a boolean are often called \textit{properties}. 
%again the same list \textcolor{red}{\texttt{w}} of durations and whose output is\footnote{In our typed language Gallina, all objects come with a type. Here, the type of \texttt{is\_leq\_10} is a function form a list of times to a list of booleans which in typed notation reads as $\texttt{is\_leq\_10} : \texttt{list} \ \texttt{time} \to \texttt{list} \ \texttt{bool}$.} a list of booleans (\texttt{true} or \texttt{false}) of the same length as \textcolor{red}{\texttt{w}}. The $i$th\footnote{We mean, the entry in the list that comes at position $i$.} element of the list \textcolor{red}{\texttt{is\_leq\_10}} of booleans is \texttt{true} if and only if the $i$the element of the list \textcolor{red}{\texttt{w}} is less than or equal (\texttt{leq}) to 10 hours. 

Finally, and similarly, we assume that we have defined/specified a function \textcolor{red}{\texttt{is\_gt\_9}} whose input is again an element of type \texttt{time} that outputs a boolean \texttt{bool} whose value is \texttt{true} just in the case that the value of the particular input is strictly more than 9.
%the list \textcolor{red}{\texttt{w}} of durations and whose output is a list of booleans (\texttt{true} or \texttt{false}) of the same length as \textcolor{red}{\texttt{w}}. The $i$th element of the list \textcolor{red}{\texttt{is\_gt\_9}} of booleans is \texttt{true} if and only if the $i$the element of the list \textcolor{red}{\texttt{w}} is greater than (\texttt{gt}) to 9 hours.

The symbols $\textcolor{red}{\texttt{\&\&}}$ is the usual boolean conjunction \emph{and}, that is, a function that has two boolean inputs and an output that is $\texttt{true}$ if and only if both inputs were $\texttt{true}$. Likewise, the three symbols $\textcolor{red}{\texttt{{=}{=}>}}$ is a single token that denotes the usual boolean implication \emph{if then}, that is, a function that has two boolean inputs and an output that is $\texttt{false}$ if and only if the first input (antecedent) were $\texttt{true}$ and the second input (consequent) $\texttt{false}$.

The functions \texttt{all} and \texttt{count} are not specified/defined by us but pertain to the standard language\footnote{The functions \texttt{all} and \texttt{count} do not exist in Rocq yet they do exist in the extension of Rocq called SSReflect. This is not relevant for the current exposition.}. The function \textcolor{red}{\texttt{all}} has as input a property and a list\footnote{The technical details to make the type of \textcolor{red}{\texttt{all}} fully precise would require some further knowledge on polymorphism that we happily shuffle under the rug here.}, and outputs a boolean \texttt{bool} that has the value \texttt{true} just in case all elements in the list satisfy that property.
%list of boolean values \texttt{v} and outputs a boolean with value \texttt{true} just in case all the elements in the list \texttt{v} have the value \texttt{true}. 
Likewise, the function \textcolor{red}{\texttt{count}} has as input a property and as output a natural number where the value of the output is equal to the amount of elements in the list that satisfy the property.
%list \texttt{v} of boolean values and outputs a natural number that equals the number of times that \texttt{true} occurs in \texttt{v}.

Finally, the symbols $\textcolor{red}{\texttt{<{=}}}$ denote the usual less than or equal (at most): in our case a function with two natural number inputs that outputs a boolean whose value is \texttt{true} exactly when the left-hand side is at most the right-hand side. Now we have explained all and every symbol in the formal language specification Art.~6.1.

Thus, \texttt{article6\_1} evaluates a given list \texttt{w} of time periods and returns \texttt{true}, {i.e.}, the driver \emph{comlies} or \texttt{false}, \emph{i.e.}, the driver does \emph{not comply}, according to the conditions defined in the function body. We have seen that the body is structured as a boolean implication, represented by \texttt{==>}, so that if the condition \texttt{is\_weeklyDP w} on the left is satisfied, then the condition \texttt{all is\_leq\_10 w \&\& count is\_gt\_9 w <= 2} on the right must also hold. Additionally, the implication's right side is a conjunction of conditions, joined by the logical \emph{and} operator (\texttt{\&\&}). In particular, \texttt{all is\_leq\_10 w} is only satisfied if every time period in \texttt{w} is less or equal to $10$. In other words, \texttt{all is\_leq\_10 w} is satisfied if and only if every element of \texttt{w} is at most $10$. Finally, \texttt{count is\_gt\_9 w <= 2} is satisfied if and only if the number of elements in \texttt{w} that are greater that $9$ is less or equal to $2$. In other words, \texttt{count is\_gt\_9 w <= 2} is satisfied if and only if there are at most two elements in \texttt{w} that are greater than $9$.

\subsection{Natural versus formal specification languages}
%In our formal language specification example we use the formal language (Gallina) of the Coq  proof assistant for various reasons. Formal languages can embody lots of different forms in mathematical expressions, however, proof assistants are designed with the purpose to check the correspondence between specifications and implementation. Therefore, while there are many different formal languages which would be equally suited for writing the formal specifications (e.g. first-order logic), the formal languages of proof assistants allow to verify the correctness of the implementations with respect to their formal specifications.
Before entering into the interactions of computational laws with our chosen legal principles, some of the core considerations for specifications in natural language versus formal language are the following.

It is widely acknowledged that the use of natural language maintains comprehensibility among all stakeholders in the legal sphere. Its broad understanding allows actors to make informed decisions about their behavior. Natural language also affords decision-making processes sufficient flexibility to interpret a given legal provision within the standing legal framework. Natural language ultimately accommodates common sense in its interpretation and application. 

Conversely, employing formal language ensures precision and eliminates ambiguity. It facilitates the application of mathematical reasoning to assess the correspondence between the specification and the resulting program. Indeed, certain types of formal specifications as already stated (such as constructive calculus, or type theory) are amenable to formal verification. Formal verification tools like the Rocq Prover or other proof-checkers provide evidence of the alignment between the specification and the code that implements it with reference to the formal ontology.

Therefore, both natural language and formal languages offer distinct advantages, albeit accompanied by potential drawbacks. Our selected example is concise and straightforward in natural language, yet it becomes highly precise and adheres to a specific format when formalized. However, it may appear complex and challenging to interpret for those unfamiliar with formal language conventions.

In Sections \ref{section:NLIssues} and \ref{sectio:FLIssues}, we dive deeper into the issues arising from our example in both natural and formal languages, examining their implications in relation to the chosen aspects of our selection of legal principles.

\section{Natural Language Issues}\label{section:NLIssues}

Art. 6.1. consists of two short lines of text with apparently straightforward meanings. In this section we will first isolate some computational complications of Art.~6.1.~when formulated in natural language. Next, we will organize the impact of such complications per legal principle. This organization will be done for general computational provisions written in natural language and for Art.~6.1.~in particular.

Since the article is so short, we repeat it here:
\begin{quote}\sf

Article 6.1: The daily driving time shall not exceed nine hours. However, the daily driving time may be extended to at most 10 hours not more than twice during the week.

\end{quote}

\subsection{Particularities of Article 6.1.~in Natural Language}

We have identified several issues with Art.~6.1.~based on natural language. 

Art.~6.1.~refers to (boldface by us):
\begin{itemize}
    \item Art.~4.~(i) \emph{‘{\bf a week}’ means the period of time between 00.00 on Monday and 24.00 on Sunday};
    \item Art.~4.~(j) \emph{`{\bf driving time}’ means the duration of driving activity recorded: \newline— automatically or semi-automatically by the recording equipment as defined in Annex I and Annex I B of Regulation (EEC) No 3821/85, or \newline— manually as required by Article 16(2) of Regulation (EEC) No 3821/85};
    \item Art.~4.~(k): \emph{‘{\bf daily driving time}’ means the total accumulated driving time between the end of one daily rest period and the beginning of the following daily rest period or between a daily rest period and a weekly rest period;}.
    
\end{itemize}

Furthermore, Art.~6.1.~remits to Art.~2 (x) of Regulation (EU) No 165/2014, of 4 February 2014, on tachographs in road transport: \emph{‘time measurement’ means a permanent digital record of the coordinated universal date and time (UTC) \footnote{UTC is the Coordinated Universal Time (UTC) considered as a worldwide time scale reference computed by the Bureau International des Poids and Mesures (BIMP). BIMP is an International Organization created by the Meter Convention in 1875 \emph{to ensure worldwide uniformity of measurements and their traceability within the international system of units} with 64 Member States and 37 associated States and economic entities.}}, 
    
From these definitions it follows that the term \emph{daily} in Art.~6.1.~becomes a technical term that may not correspond to a traditional 24-hour period\footnote{For instance, \emph{day} is not defined in the Regulation nor in its framework, although \emph{daily} is. Also, the regulation uses 24-hour periods to decide upon the compliance with its articles. If we follow Art.~4 (j) very strictly it seems that no daily driving is possible between two consecutive weekly rest periods, or when a worker starts his/her profession (and thus not had a rest period ever before).}. We deem this undesirable since it can easily be misleading to users and coders of the Regulation alike. 

The opposite is the case of the term \emph{week}, which is explicitly defined in Regulation 561/2006 as \emph{the period of time between 00.00 on Monday and 24.00 on Sunday}\footnote{We stress that 00:00 and 24:00 may mean the exact same moment.}, and corresponds to the common use of \emph{calendar week} in everyday language. Such issues, although also present in the law when applied in an analog (traditional or non automated) manner, are prone to be more misleading for coders, who necessarily should make an interpretation decision when implementing the \emph{computational} provisions using or related to such terms.

Another issue involves \emph{leap seconds}\footnote{For the meaning and importance of leap seconds see, e.g. NIST's Leap Seconds FAQs at the following link \url{https://www.nist.gov/pml/time-and-frequency-division/leap-seconds-faqs}}. The regulation specifies what constitutes a \emph{week} and adopts a calendar week, but it does not define how leap seconds should be treated. Suppose that there is a negative leap second on Sunday. Thus, this particular Sunday has ended at 23:59:59 instead of 24:00:00 and the moment 24:00:00, will not exist for this particular Sunday\footnote{Actually, strictly speaking the moment 24:00:00 does not exist as a legitimate time notation. By tacit convention, it is often seen as an alternative notation for 00:00:00 the next day.}. How to interpret \textit{week} as per Art.~4 (i)? Should it now span a period of \emph{14 days} so that it goes to the next moment where Sunday 24:00 does exist? Although this is not according to the purpose of the law, it is not to the coder to conclude this\footnote{We are considering that the hours referred to - 9 hours or 10 hours - are \emph{durations} by the wording of Art.~4 (k) as \emph{accumulated driving time}. While some \textit{hours} may be affected by leap seconds, duration hours are not.}. We are of the opinion that it was not the intention of the legislator to allow for different coders to freely interpret this term.

However, the most serious problem concerning Art.~6.1.~and natural language specifications highlighted by our example is that daily driving times spanning two weeks (so, starting on a Sunday and ending on a Monday) can be attributed to either week. Legal compliance with the maximum daily driving times of one week will eventually depend on the attribution of such periods in future weeks\footnote{See \cite{LeNozzediGiustizia} for details.}.

\subsection{Legal Principles in Natural Language Law}

Our example reflects how legal provisions are currently written in natural language, regardless of whether they are intended to be enforced by human agents or ADM\footnote{We are not entering into discussions around the ongoing efforts on the “Code as Law”, for example, in Denmark \url{https://en.digst.dk}. Nor do we consider structured natural language for law as in the \href{https://github.com/CatalaLang/catala}{Catala programming language}, or in NASA’s \href{https://software.nasa.gov/software/ARC-18066-1}{FRET framework}. Neither do we consider other issues such as correspondence between updates in Law and software applying that law via ADM or viceversa.}. These provisions  often serve as the sole specifications provided by legal instruments when they are to be applied by a software system as in the case of a computational law.

\subsubsection{Legality} \label{subsubsection:NL:legality}
The Principle of Legality and natural language specifications for ADM pose the initial query on the type and procedures enabling the implementing software. If the decision-making process is fully automated, the ADM software may act as a legal agent. This raises questions, for example, whether prior legal authorization is required to implement the autonomous ADM system or which Public Administration body is responsible for the ADM decisions.

Natural language specifications are produced by the legitimate actors following legally established processes and are accessible to the society. However, as seen in Art.~6.1. they may leave room for interpretation. If that is the case, who and under which procedures should be entitled to make the choice of the meaning to be implemented by the ADM software and how should that choice be publicized? 

Furthermore, if the implementation is outsourced to private actors, does the interpretation correspond to the public body? Or is it left to the private enterprise in charge of the implementation? Should legal procedures and guarantees be established prior to such outsourcing? As shown by our case study, this is not currently the case.

As showcased by our example, natural language specifications, due to vagueness, underspecification, or ambiguity, can generate \emph{erroneous}  implementation due to the interpretation made on the basis of the natural language specifications\footnote{We emphasize the word \emph{erroneous} since often being an error or not depend on interpretation choices. The Research Project, led by Dr. Jordi Ferrer Beltran and Dr. Diego M. Papayannis (University of Girona, Los errores en la producción y en la aplicación del derecho (EPAD) (PID2020-114765GB-I00) has dealt with the meaning and impact of errors in legal proof reasoning in Procedural Law.}. Moreover, if the ADM acts as an autonomous legal agent, an ADM system tailored to natural language specifications can be more prone to errors.

However, natural language fits traditional legal practice, responds to the Principle of Legality, and allows easier observation of the legal framework\footnote{We do not enter to consider liability issues, since liability fall outside of the limited scope of this paper.}.

\subsubsection{Legal Certainty} \label{subsubsection:NL:legalcertainty}

We have seen above that Art.~6.1.~when formulated in natural language is actually ambiguous. Consequently, Legal Certainty can be compromised at the level of foreseeability, stability, and consistency. Thus, to implement Art.~6.1.~one first needs to disambiguate. Therefore, interpretations/disambiguations of this provision by non-experts may compromise the \emph{foreseeability}, as each programmer may interpret and implement a legal provision slightly differently. Consequently, \emph{stability} will also be affected as exemplified by Art.~6.1.~’s oversight of less frequent but periodic issues such as \emph{leap seconds}. \emph{Consistency} will suffer both if different interpretations and implementations are made and if disparate cases are treated uniformly. 

Legal stipulations, however, are required to be duly published to satisfy the Principle of Legal Certainty. Since the text of the legal instrument usually comprises the natural language specifications or instructions, its accessibility on official, public platforms is ensured, reaching all actors uniformly, even though, \emph{computational} provisions may be implemented in a way that results in a non-uniform manner.

The positive aspect of \emph{computational} laws written in natural language, however, is that such language is naturally embedded in institutions and, specially, in the citizenry framework due to the long tradition of analog laws.

\subsubsection{Prohibition of Arbitrariness} \label{subsubsection:NL:arbitrariness}

As seen in the previous subsection, coders often rely solely on the text of the legislative acts as instructions for designing ADM software and, therefore  may be in a position of interpreting provisions such as Art.~6.1. Coders' interpretations, regardless of how well-intentioned, apparently reasonable, proportionate, and unbiased they may be, are merely one among several possible interpretations. Opting for a specific interpretation of a legal provision without justifying the reasons for the choice may later render the ADM decision arbitrary and the responsible Public Administration liable.

To avoid arbitrariness in ADM, international organizations and domestic strategies are pushing for the ethical principle of explainability of ADM software programs\footnote{One such example is the EU AI Act. For instance, the Council of Europe, the OECD, the UNESCO are considering \emph{explainability} as one of the main characteristics a human-centered AI should display.}. Although the programmers' choices may appear explicit if the software program complies with an \emph{ex ante} requirement of explainability, merely stating how the ADM software program operates does not reveal why the coder interpreted the legal text in a particular way, nor does it ensure consideration of all possible legitimate interests.

Additionally, scholars are raising concerns about the legitimacy of programmers interpreting and actively choosing how to implement legal texts\cite{Godfrey03052024}. This concern is even more pronounced when a private third party develops the ADM software program, as conflicts of interest may remain unnoticed. Intellectual property rights over ADM software programs pose significant challenges for third-party software acquired through public procurement and for in-house administrative ADM software\footnote{One example of the difficulties of accessing the code of in-house administrative ADM software is presented by the ongoing Spanish case on the BOSCO algorithm. For the discussion on the grounds stated by the plantiff, Fundación Civio, and the judgements, see Huergo, A. \emph{Por qué aciertan las sentencias sobre el ‘algoritmo’ del bono social eléctrico}, and Ponce, J. \emph{Por qué se equivocan las sentencias sobre el algoritmo del bono social eléctrico}, both of 2024 and available at the Blog Almacén de Derecho \href{https://almacendederecho.org}{https://almacendederecho.org}. For the ulterior conversation, see \emph{¿Utopía o distopía algorítmica?} entry of the Blog Nudging aplicado a la Mejora de la Regulación y al Uso de Algoritmos y de Inteligencia Artificial. \href{https://rednmr.wordpress.com/2024/05/17/utopia-o-distopia-algoritmica/}{https://rednmr.wordpress.com/2024/05/17/utopia-o-distopia-algoritmica/} For the BOSCO case, \emph{vid.} Ponce, J., The energy social bonus and the Bosco program: about algorithms, bugs and source code. Regarding the first court decision handed down in 2021: a bad judgment that we hope will be corrected soon \url{https://rednmr.wordpress.com/2022/09/29/the-energy-social-bonus-and-the-bosco-program-about-algorithms-bugs-and-source-code-regarding-the-first-court-decision-handed-down-in-2021-a-bad-judgment-that-we-hope-will-be-corrected-soon/}}.

Moreover, ADM programs will inevitably produce undesirable outputs as decisions at times. Those undesirable outputs may not follow common sense or analogical application, rendering them arbitrary by default. Any deviation from the legislative intent that results in a decision with legal effects makes that decision unjustifiable from the perspective of the concerned individual\footnote{Different legal systems have slightly different criteria as to determine the purpose of the laws. Though, commonly, the explanatory memorandum of the legal instrument will express explicitly the goal and basic functioning of the law in question.}.

While decision-making authorities may also encounter problems of undesired outcomes from ADM software outsourced to third parties, the Administrative Unit signing the ADM decision provides legal coverage for the ADM software program. These authorities are better positioned to access and process the data and software logic leading to the outcome. However, this privileged position does not necessarily ensure a fair decision\footnote{We use \emph{fair} here as the antonym of \emph{arbitrary}. We do not engage in debates around the \emph{fairness of algorithms} and similar issues in this paper.}. It is imperative for regulators to recognize that ambiguities, underspecifications, and contradictions in the legal text and the broader legal framework can impede impartial ADM decisions.

In the context of our example, Art.~6.1., beyond selecting a specific natural language interpretation, the ability to choose whether to assign the permitted extended driving periods to one week or the next could lead to arbitrariness. The regulation's disregard of leap seconds and their effect on the allocation of extended driving periods could result in designated public authorities sanctioning cases that should not be fined according to the regulation. This could also lead to inconsistent treatment of essentially similar cases.

\subsubsection{Motivation} \label{subsubsection:NL:motivation}

The key issues concerning natural language specifications and the possibility of contesting or challenging an ADM decision implemented into software lie in being aware that a given decision has been automated,  having the opportunity to access a contradictory process to challenge the ADM software's decisions, and fundamentally, being granted a fully reasoned decision that clearly displays the legal and factual grounds underpinning such a decision. 

Automated decisions typically do not provide sufficient reasoning on the rationale of the decision-making process. Consequently, the affected individual may encounter difficulties in effectively challenging an automated decision, as the possibility of challenging an ADM decision before it is issued must be intentionally and proactively integrated into the decision-making process. This usually necessitates a non-automated actor to study the variety of allegations made on a previously duly reasoned decision.

On the other hand, authorities may find it problematic to explain the functioning of the software, even when its specifications are contained in the legal text, as natural language provisions. Although it may seem relatively easy to craft allegations, given that the software instructions are accessible and understandable to anyone, ambiguities and imprecisions make it challenging to comprehend how the ADM software was tailored. Even if the problems of legal reasoning are overcome, affected individuals usually have to recall large amounts of data to prove their point\footnote{In the case of Regulation 561/2006, such data will either be automatically recorded by a tachograph or manually introduced in the corresponding cards, or in both media. Arts.~6 and 8 of Regulation 561/2006, and Arts.~29 and 30 of Regulation 165/2014, detail these recording requirements, which further complicate the challenge process due to the volume and nature of the data involved.}.

The lack of transparency in ADM software and the absence of motivation constitute significant obstacles for individuals attempting to precisely detect and prove the incorrect selection of data or its flawed processing by the decision-maker. This opacity complicates the task of holding the deploying authority accountable, especially when the burden of proof regarding the malfunctioning or unfairness of the decision rests on the affected individual.

In the case of Art.~6.1.~, the tension between the Principle of Motivation and the overall difficulty of pinning down what constitutes a \emph{daily driving period} further exacerbates these challenges. Any ambiguity or underspecification in defining critical terms undermines the individual's ability to effectively challenge and seek redress for potentially erroneous automated decisions.

On the other hand, natural language specifications can be used by the authoring institution to craft a motivation of the decision made, resorting, for instance, to the wording of the article of our example and connecting it to the legal frameworks in which it is inserted. However, such motivation does not necessarily connect with the real functioning of the software. \emph{Mutatis mutandis} to crafting of a motivation disconnected from the actual functioning of the ADM system can be said about the affected individual who attempts to challenge the ADM decision through arguments based solely on the natural language legal provisions. The affected individual will not know how the ADM system works in reality. Instead, s/he will have to guess the software's functioning based on the text of the legal provision executed by the ADM system.

\subsubsection{Clarity} \label{subsubsection:NL:clarity}

The Principle of Clarity mandates that regulations and decisions emerging from the executive branch should be understandable to the ordinary citizen. Citizens should easily understand natural language specifications,  especially if the legislator has followed good regulation practices and observed the Principle of Clarity. 

Art. 6.1. seeks to establish how to check compliance with daily driving periods. However, the legislator has not achieved its goal given ADM coding requirements. The coders of the ADM system will have to make a choice. Only they will know exactly how they implemented the provision and why. This would lead to the software probably constituting a \textit{black box }for the non-expert deployers and the citizens. It would also complicate the task of future lawyers and judges examining the rationale and, subsequently, also the legality of a sanction. Such \textit{opaqueness }collides with the need for \textit{transparency }and \textit{motivation }of administrative decisions.

\section{Formal Language issues}\label{sectio:FLIssues}

An important feature of formal languages is that they allow for so-called formal methods. Formal verification is the most important in the context of this paper so we start by discussing this.

\subsection{Formal Verification}\label{section:formalVerification}

We recall that the A in ADM stands for `Automated'. However, virtually all automation will contain errors. Such an error can be a typo from the programmer or a slight reasoning error encoded in the computer program that automates the decision-making. These errors can occur both in the scenario when specifications are issued in natural language as in Subsection \ref{section:natLangSpecArt61} and in the scenario when specifications are issued in formal language as in Subsection \ref{section:formLangSpecArt61}. However, when we are to write computational laws, formal languages have one major benefit over natural languages: formal methods. 

Formal methods allow us to prove correctness of code with respect to the specification mathematically. We cannot stress enough that mathematically proving the correctness of a program cannot be done if the specification of that program is written in natural language. What happens if our proof of program correctness is wrong? Typically, ADM programs contain thousands of lines of code. Consequently, the proofs of correctness will become very large. It seems we have only moved faith in the correctness of the ADM program to faith in the correctness of our proof\footnote{Due to the nature of proofs, it seems harder though to make an error in a correctness proof than in the code. We cannot really sustain this informal claim that mostly rests on experience.}. This is where so-called \emph{proof assistants} enter the stage. 

Proof assistants are small computer programs that check the proof of correctness of the code with respect to its specification in formal language. So now we have shifted faith in the ADM code to faith in the proof assistant. And actually, this is right. However, we did gain much along the way. Firstly, coming up with a proof of correctness of ADM software is a genuine intellectual act that requires due analysis of the ADM software. Typically, at this stage, many bugs are found in the software. Secondly, when applying the proof checker to our alleged proof, we trust once a small program: the proof assistant. The exact proof assistant can be applied repeatedly to various correctness proofs of various ADM software. Instead of an act of faith in the correctness of every ADM software, we now have one act of faith about the correctness of the proof assistant: a relatively small program that stays quite stable over time and has been debugged intensively over decades. 

In this paper, we have chosen the formal specification language Gallina since it is expressive enough: it can naturally talk about various ontologies from different computational laws. Also, Gallina can be used to reason\footnote{The proof assistant that corresponds to Gallina is called Rocq \emph{vid.} footnote 19.} about program behavior and, as such, can serve the purpose of an \emph{understandable} formal specification language that determines the nature of a computational law so that the same specification can be used to prove the correctness of a program that implements the specification. 

Employing proof assistants to prove program correctness yields (as good as) zero-error software. It is important to note that the resulting software will be as good as the corresponding formal specification. The problem of whether the formal specification fully aligns with our intention cannot be solved by mathematical means. Between intention and a linguistic materialization of this intention, there is always a translation at work that will introduce uncertainty. Something similar occurs with law written in natural language: we do not know with certainty that a law fully reflects our ethical intuitions and conception of fairness\footnote{However, traditional law-making has put into place some guardrails to ensure that the law is as much as possible in line with the intention behind it through good practices, public consultations, and impact assessments. As for the EU, Better Regulation toolbox is a relevant initiative \url{https://commission.europa.eu/law/law-making-process/planning-and-proposing-law/better-regulation_en}}. 

\subsection{Legal Principles in Formal Language Law}

We will now analyze benefits and drawbacks of formulating computational law in a formal language. We will use Art.~6.1.~as a mere illustration in this analysis.

\subsubsection{Legality} \label{Subsection:NF:Legality}

If we consider Legality an integral part of the Supremacy of the Law, the initial query arising from issuing specifications in formal language is who holds the rightful authority to issue the specifications. Is it the legislative or the executive branch that should be in charge of the  development of the legal instruments? 

Another question concerns public access to both the specification and the corresponding implementation.
The formal specifications should be made public as any other legal instrument. However, the source code of ADM software is typically not open-source neither is it published on an official public platform with institutional guarantees\footnote{An exception can be found in French Administrative System in the CodeGouv project: \url{https://code.gouv.fr/sources/}. As per 2021, source code used by the French Administration should be made public. See also \cite{Guadagnoli:2021:OpenSourceFrance}.}. Consequently, an accessibility issue for proof-checking the correctness of the ADM software to the formal specification compounds on a comprehensibility problem associated with formal specifications and Legality\footnote{Access to formal specifications is generally provided by Transparency Bodies and the Judiciary across the EU, while access to source code, whether in-house or third-party, remains a more contentious matter. Even if full access to all information regarding ADM software is permitted ---it is usually granted upon request--- thereby imposing an additional formal step on the public to understand how they should act or to challenge an automated decision after it has been issued, serving as an \emph{ex post} mechanism.}.

As shown by our example in Subsection \ref{section:formLangSpecArt61}, Art.~6.1.~written in Gallina, is not easily understandable. However, with formal specifications emerging directly from the legislative or the executive, the Public Administration tasked with verifying compliance is certain about the legality and univocal meaning of such formal specifications. 

Regarding Art.~6.1., the Public Administration is also operating under the assumption that the ADM system is accurate. The availability of formal specifications issued through legislative or regulatory processes can be presumed to generate software that is by and large correct to the specifications, regardless of whether the software implementation is in-house or entrusted to private entities. In the case of Art.~6.1. the provision is enforced through different private software acquired and used by different authority bodies (mainly law enforcement) tasked with verifying compliance with the road transport regulatory framework within the European Economic Area. Democratically issued formal specifications could ensure that their implementation is uniform. The resulting ADM software programs would also be susceptible to formal verification.

That being said, the Principle of Legality faces challenges, especially when legislation written in formal language is incomprehensible to both the general public and the institutions tasked with applying the law. It is a reality that the majority of the population, being street-level civil servants or the citizenry, is not proficient in coding or formal languages.

\subsubsection{Legal Certainty} \label{Subsection:NF:LegalCertainty}

Let us now see how the Principle of Legal Certainty fares when computational law is written in a formal language. The main problem remains the incomprehensibility of formal language for both the non-versed executive civil servants and the citizenry. As seen in Subsection \ref{section:formLangSpecArt61}, the possible formal specification of Art.~6.1.~can be difficult to understand. It requires a decent level of education in mathematical notation. Furthermore, as already stressed, the formal language lines corresponding to Art.~6.1.~are not the entire specification. For instance, localizing and understanding the \emph{type} definitions specified at earlier stages add to general public's difficulties will face with understanding and accessing formal language computational laws.

A specification in formal language is precise and unambiguous, helping explain the software's functioning. Legislators and public administrations can harness this capability to certify software quality. Even if the underlying mathematical proof is not comprehensible to the general public, public certification will help citizens and legal professionals identify the expected operation of ADM systems and simplify redress. 

\subsubsection{Prohibition of Arbitrariness} \label{Subsection:NF:Arbitrariness}

Deploying a formal specification ensures that the ADM system is fully deterministic, issuing the same output for the same input every time. As we have seen for Art.~6.1., formalized as a function, there is no room for underspecification or ambiguities. Each part of the function is explicitly stated or leads to an element that is explicitly stated elsewhere, as for example, the function \texttt{is\_weeklyDP}. This uniformity and unambiguity eliminates the risk of differing interpretation. 

The deployment of formal language for software specifications ensures that the resulting software is absolutely precise if proof assistants are used to establish exact correspondence between specification and code. Consequently, the resulting ADM software is impartial. 

\subsubsection{Motivation} \label{Subsection:NF:Motivation}

As pointed out in Subsubsection \ref{subsubsection:NL:motivation}, ADM systems often fail to provide sufficient motivation for their decisions. While formal specifications can enhance \emph{explainability} by detailing what the software does and how it functions, these measures alone cannot fully address the absence of motivation. Explainability is fundamentally descriptive focusing on the software's functioning, but it does not provide insight into why a particular decision was reached in a specific context. Motivation, by contrast, seems to be inherently constructive. Motivation involves justifying the decision-making process by articulating the reasoning based on the relevant facts and the legal grounds of the specific decision made.

However, explainability provided by the formal specifications, on the one hand, and the possibility of using formal methods to verify the correctness of the software with respect to the formal specifications, on the other hand, can mitigate the lack of motivation by ensuring the individual that the ADM system is functioning as the liable administration intended, thus facilitating redress.

\subsubsection{Clarity} \label{Subsection:NF:Clarity}

As for the interaction between the Principle of Clarity and the formal specifications the main issue is again understandability. While comprehensibility issues for the Principle of Legal Certainty accompany general access and publicity, the Principle of Clarity operates within the context of the administrative procedure.  The administrative procedure of decision-making is overall difficult for both the authority bodies and the affected citizens if formal language specifications are used. As for Art. 6.1., for example, both the citizenry and the civil servants would need to understand that the data is stored in  $w$, what elements $w$ comprises (durations), if their format satisfies $is\_weeklyDP$, etc.
 
We also draw attention to the question of whether the Principle of Clarity should apply to the data to be retrieved for the purpose of evaluating or reevaluating whether a given behavior, specifically in the context of driving activity, is sanctionable. For the non-expert deployer and for the citizen to access and read the data leading to the outcome can be very difficult, since its format and quantity can be unmanageable for the humans or it would require an amount of time superior to the legally established deadlines. Clarity for the data handled in a given decision highlights the importance of transparent and understandable data-handling practices, which are essential for ensuring fairness and accountability in administrative decision-making processes.

\section{Conclusions}\label{section:Conclusions}

The push for greater efficiency, effectiveness, and timeliness in public administration, coupled with the growing volume of data and the increasing complexity of legal systems, have been key driving forces behind the digitization of public administrations. ADM systems are being progressively implemented, not only for predictive analysis and the formulation of public policies but also for the application and verification of compliance with existing legal instruments. 

This shift has led to a growing body of computational laws or legal provisions designed to be executed and enforced through ADM systems. Such laws should be drafted to facilitate their seamless integration into these systems. However, we have seen that natural language computational laws present inherent difficulties, while formal specifications used in ADM software come with their own set of challenges.

The table below summarizes our main findings. Per legal principle, we have organized the assessment of using one language or another for the specification of legal ADM. 

While natural language specifications fit traditional legal practice, their potential for underspecification and ambiguities are important drawbacks.

And while formal specifications are obscure and complex for the non-experts, they eliminate ambiguities and underspecifications and, allow for the deployment of formal verification of specifications and software correspondence.

At this stage in the evolution of ADM systems, society must democratically determine the appropriate balance between natural and formal language in drafting of \emph{computational} laws, taking into account the trade-offs between the enhancement of certain legal principles over others. It may well be that both languages can be combined or that so-called Structured Natural Languages (see, e.g.~\cite{HuttnerMerigoux:2022:CatalaFuture, Giannakopoulou2020FormalRE}) can be part of legal ADM practice in a near future. These options have not been considered in this paper. As a first step, we investigated the impact of two scenarios: natural language or formal language for legal ADM specifications.

\begin{center}
\begin{table}
\begin{tabular}{|l|p{0.4\textwidth}|p{0.4\textwidth}|}
 \hline
\multicolumn{3}{|c|}{Conclusions Table}\\
\hline
% Art. 6.1. 
 & Natural Language 
 & Formal Language \\ \hline
 %%%%%%%%%%%%%%%%%%%%%%%%%%%%%%%%%%%%%%%%%%%%%%%%
 %%%%%%%%%%%%%%%%%%%%%%%%%%%%%%%%%%%%%%%%%%%%%%%%
 Legality 
 & Subsubsection \ref{subsubsection:NL:legality}
 \newline - SALA may necessitate a prior legal authorization; \newline - possible externalization of administrative ADM without pre-established legal procedures and guarantees;  \newline
 - more prone to generating erroneous software\newline
 + fits traditional legal practice
 & Subsubsection \ref{Subsection:NF:Legality}; \newline 
 %- possible lack of legitimacy of the actors issuing the specifications in the formal language;  \newline - possible lack of transparency and of accountability \newline
 - current legal eco-system largely incompatible \newline
 + software more likely to be according to specification
\newline
+ allows for formal methods certifying correctness \\ \hline  
%%%%%%%%%%%%%%%%%%%%%%%%%%%%%%%%%%%%%%%%%%%%%%%%
 %%%%%%%%%%%%%%%%%%%%%%%%%%%%%%%%%%%%%%%%%%%%%%%%
Legal Certainty   
&Subsubsection \ref{subsubsection:NL:legalcertainty} \newline - difference between \emph{daily-life} and legal definitions of the terms\newline
- underspecification of the terms, leaving wide margins of interpretability\newline
+ naturally embeds in citizen's reference frame
&Subsubsection \ref{Subsection:NF:LegalCertainty}  \newline - incomprehensibility of the formal language for the citizenry \newline - explainability of the software does not necessarily entail the ability to accurately predict its behavior\newline
+ no underspecification nor ambiguity\\ \hline
%%%%%%%%%%%%%%%%%%%%%%%%%%%%%%%%%%%%%%%%%%%%%%%%
 %%%%%%%%%%%%%%%%%%%%%%%%%%%%%%%%%%%%%%%%%%%%%%%%
Arbitrariness
 & Subsubsection \ref{subsubsection:NL:arbitrariness} \newline - ambiguities and contradictions are resolved without further justifying the option taken as to the meaning of the ambiguous term
 & Subsubsection \ref{Subsection:NF:Arbitrariness}  \newline+ fully deterministic \\  \hline
 %%%%%%%%%%%%%%%%%%%%%%%%%%%%%%%%%%%%%%%%%%%%%%%%
 %%%%%%%%%%%%%%%%%%%%%%%%%%%%%%%%%%%%%%%%%%%%%%%% 
 Motivation
 & Subsubsection \ref{subsubsection:NL:motivation} \newline 
 %-  not provided by mere explainability of the ADM system \newline 
%- opacity hampers possibilities of legally challenging the ADM outcome
+ natural language can be used to provide a motivation; \newline
- motivation not necessarily ties up with the functioning of the automated implemented reasoning
& Subsubsection \ref{Subsection:NF:Motivation} \newline - execution of a software does not provide an explanation; \newline + proof with formal methods of correctness can partly mitigate lack of motivation; 
%\newline + proof the software's correctness adds to the requirement of due dilligence 
\\
 \hline
%%%%%%%%%%%%%%%%%%%%%%%%%%%%%%%%%%%%%%%%%%%%%%%%
 %%%%%%%%%%%%%%%%%%%%%%%%%%%%%%%%%%%%%%%%%%%%%%%%
 Clarity 
  & Subsubsection \ref{subsubsection:NL:clarity} \newline + easy to understand by the citizenry; \newline -SALA may constitute a black box 
 & Subsubsection \ref{Subsection:NF:Clarity} \newline - inintelligibility of the regulation for most citizens \newline - difficulties in retrieving the data driving the outcome \\
\hline
\end{tabular}
\caption{Abbreviation: SALA stands for Software Acting as Legal Agent. If there is a plus sign upfront the observation is considered a positive aspect. Likewise, negative aspects are preceded by a minus sign.}
\end{table}
\end{center}

\bibliographystyle{alpha}
\bibliography{biblio2.bib}

%\nocite{*}
%\printbibliography

%\bibliographystyle{plain}
%\bibliography{biblpl}
%\bibliography{biblpl}
%\bibliographystyle{apa}
\end{document}